\documentclass[prc,reprint,onecolumn,amsmath,amssymb,superscriptaddress]{revtex4-1}
\usepackage{graphicx}
\newcommand{\be}{\begin{equation}}
\newcommand{\ee}{\end{equation}}

\begin{document}

\title{Legendre Analysis of Differential
	Distributions \\ in Hadronic Reactions}

\newcommand*{\PNPI}{Petersburg Nuclear Physics Institute, NRC 
	Kurchatov Institute, Gatchina, 188300, Russia}
\newcommand*{\GWU}{The George Washington University, Washington, 
	DC 20052, USA}

\author{Yakov~I.~Azimov}
\thanks{Deceased.}
\affiliation{\PNPI}
\author {Igor~I.~Strakovsky}
\email{igor@gwu.edu}
\thanks{Corresponding author.}
\affiliation{\GWU}
\author{William~J.~Briscoe}
\affiliation{\GWU}
\author{Ron~L.~Workman}
\affiliation{\GWU}

\date{\today}


\begin{abstract}
\centerline{\large Abstract}
\vskip 0.15in
Modern experimental facilities, such as CBELSA, ELPH, JLab, MAMI and SPring-8 
have provided a tremendous volume of data, often with wide energy and angular
coverage, and with increasing precision. For reactions with two hadrons in the final
state, these data are often presented as multiple sets of panels, with
angular distributions at numerous specific energies. Such presentations have
limited visual appeal, and their physical content is typically extracted through some 
model-dependent treatment. Instead, we explore the use of a
Legendre series expansion with a relatively small number of essential coefficients. 
This approach has been applied in several recent experimental investigations.  
We present some general properties 
of the Legendre coefficients in the helicity framework and consider what
physical information can be extracted without any model-dependent assumptions.
\end{abstract}


\maketitle

\section{Introduction}
\label{sec:intro}

Modern detectors, combined with the present generation of 
accelerator facilities, are capable of
providing large reaction-specific sets of experimental data. 
These sets have often been combined in partial-wave analyses
with the hope of extracting elements of the fundamental reaction
process (such as resonance parameters and coupling constants).
The analyses generally have some model-dependence and are limited
by the quality of the available data. 

Here we address the problem of displaying these large data sets,
evaluating their physical content, and determining their sensitivity 
to partial-wave content in a model-independent manner.  
Even in relatively
simple cases of $2\to2$ reactions, data are usually presented
as multi-panel pictures with a great number of angular distributions
for different energies (and/or energy distributions for different angles).
Such an approach can be used, of course, to compare with various models,
but is not practical for any direct extraction of physical information.

In several recent works, we and others have suggested and applied another approach,
involving the expansion of differential cross sections, for both
unpolarized~\cite{CBC,A2} and polarized~\cite{CLAS} photoproduction
of single pseudoscalar mesons, in terms of Legendre coefficients. For
a limited energy interval, it appears sufficient to use a finite 
number of the expansion terms, which may be plotted as a function of energy, 
thus providing a more clear and visually suggestive
presentation, which may be further analyzed through models or partial-wave
analyses.

Preliminary results of this study were reported at the \textit{Hadron 
Structure and QCD: from Low to High Energies} Workshop~\cite{YaA16}. 
In the present paper, we further describe and study this 
approach more systematically.  Then we discuss its utility for the
extraction of model-independent information.

\section{Expansion of Amplitudes and Cross Sections}
\label{sec:expan}

Let us consider $2\to2$ reaction
\be
	a+b\to c+d\,,
	\label{reac}
\ee
where particles have spins $s_a, s_b, s_c, s_d$. It can be described
by helicity amplitudes~\cite{jw}
	$A_{\lambda_c\lambda_d;\,\lambda_a\lambda_b}(W,z,\phi)$,
where $\lambda_a,\,\lambda_b,\,\lambda_c,$ and $\lambda_d$ are the
corresponding $s$-channel helicities, $W$ is the center-of-mass (c.m.) 
energy, and $z=\cos\theta\,$. The angles $\theta$ and $\phi$ are, respectively, 
the polar and azimuthal c.m. angles. These amplitudes may be decomposed 
in terms of the Wigner harmonics
\begin{align}
	A_{\lambda_c\lambda_d;\,\lambda_a\lambda_b}(W,z,\phi) &=
	F_{\lambda_c\lambda_d;\,\lambda_a\lambda_b}(W,z)\,
	e^{i(\lambda-\mu)\phi}\, \\
 &= \sum_j(2j+1)\,f^j_{\lambda_c\lambda_d;\,
	\lambda_a\lambda_b}(W)\,d^{\,j}_{\lambda\mu}(z)\,
	e^{i(\lambda-\mu)\phi}\,
	\label{h_amp}
\end{align}
with partial-wave amplitudes
\begin{align}
	f^j_{\lambda_c\lambda_d;\,\lambda_a\lambda_b}(W)=\frac{1}{ p}\,
	\langle\lambda_c\lambda_d|T^j(W)|\lambda_a\lambda_b\rangle~,
\end{align}
being elements of the $T$-matrix, related to the $S$-matrix,
$T=(S-1)/(2i)$, $p$ being the initial relative c.m. momentum, and
\be
	\lambda=\lambda_a-
	\lambda_b\,,~~~~\mu=\lambda_c-\lambda_d\,.
	\label{lamu}
\ee
The scattering/production angle $\theta$ is taken to be the angle
between the c.m. momenta of particles $c$ and $a$ (or $d$ and $b$). 
All the values of $j,\lambda,\mu$ are simultaneously either integer 
or half-integer and the above $j$-summation runs over all physical values of
$j\geq|\lambda|,|\mu|$.

The differential cross section, with all
initial and final helicities fixed, is~\cite{jw}
\be
	d\sigma_{\lambda_c\lambda_d;\,\lambda_a\lambda_b}=
	\left(\frac{2\pi}{p}\right)^{\!2} \left|A_{\lambda_c\lambda_d;\,
	\lambda_a\lambda_b}(W,z,\phi)\right|^2\,d\Omega\,.
	\label{crsec}
\ee
Note that $d\sigma/d\Omega$ is independent of the azimuthal angle
$\phi\,$, since every particular helicity amplitude depends on $\phi$
only through a phase factor.

The totally unpolarized differential cross section can be written as
\begin{align}
	\frac{d\sigma(W,z)}{d z} &= N\sum_{(\lambda, \mu)}\,
	|F_{\lambda_c\lambda_d;\,\lambda_a\lambda_b}(W,z)|^2 \\
	&= N\sum_{(\lambda, \mu)}\, \sum_{j_1,j_2}\,(2j_1+1)
	(2j_2+1)\,f^{j_1\,*}_{\lambda_c\lambda_d;\,\lambda_a\lambda_b}(W)\,
	f^{j_2}_{\lambda_c\lambda_d;\,\lambda_a\lambda_b}(W)\,
	d^{\,j_1}_{\lambda\mu}(z)\,d^{\,j_2}_{\lambda\mu}(z)\,,
	\label{unpcrsec}
\end{align}
where $\sum_{(\lambda, \mu)}$ implies summation over all initial and final
helicities. We separate out the normalization factor $N$, which has a simple
structure, with a known dependence on the energy and on the
spins (due to summing and averaging over polarization states), but is
quite independent of any dynamics, helicities, angles, and angular momenta
(it is not essential for the following discussion). The angular
dependence of each summand in Eq.(\ref{unpcrsec}) is completely described
by two $d$-harmonics with the same $\lambda$ and $\mu\,$. Their product 
can be decomposed into a Clebsch-Gordan series over the Legendre 
polynomials.  As a result, we obtain
\be
	\frac{d\sigma(W,z)}{d z}=\sum_{J=0}^{\infty}\,A^{(\sigma)}_J(W)\,P_J(z)
	\label{decomp}
\ee
with integer $J\,$. According to the composition rules for angular momenta, 
every $A^{(\sigma)}_J(W)$ contains bilinear contributions of partial-wave 
amplitudes (see Eq.(\ref{unpcrsec})) with angular momenta $j_1$ and $j_2$ 
satisfying the familiar relations
\be
	|j_1-j_2|\leq J \leq j_1+j_2\,.
	\label{3j}
\ee
This means that a particular Legendre coefficient $A^{(\sigma)}_J(W)$ 
generally contains an infinite number of contributions from partial-wave amplitudes 
with various $j$-values. But it evidently can not contain interference of 
amplitudes with too different $j_1$ and $j_2$ ({\it i.e.}, with $|j_1-j_2|>J$). 
Quadratic terms, having $j_1=j_2=j\,$, may appear only at sufficiently large  
$j\geq J/2$. Of course, the coefficient $A^{(\sigma)}_0(W)$ coincides with half 
the total cross section of reaction (\ref{reac}) at the \mbox{energy $W$}. 
Recall that $\sigma^{\rm tot} = 4\pi A^{(\sigma)}_0$ if one fits 
$d\sigma /d\Omega$.

The Legendre coefficients have another, less evident, property. To derive it, we
combine Eqs.(A1), (A2), and (41) of Ref.~\cite{jw} and obtain
\be
	d^{\,j}_{\lambda\mu}(-z)\,\langle\lambda_c\lambda_d|T^j(W)\,|\lambda_a
	\lambda_b\rangle =\frac{(-1)^{s_a+s_b-\mu}}{\eta_a \eta_b}\, d^{\,j}_{\lambda'\mu}(z)
	\,\langle\lambda_c\lambda_d|T^j(W)\,P|\lambda'_a\lambda'_b\rangle\,,
	\label{-z}
\ee
\be
	F_{\lambda_c\lambda_d;\,\lambda_a\lambda_b}(W,-z)=\frac{(-1)^{s_a+s_b-\mu}}
	{\eta_a \eta_b}\,F^{(P)}_{\lambda_c\lambda_d;\,\lambda'_a\lambda'_b}(W,z)\,.
	\label{F-z}
\ee
Here $\lambda'_a=-\lambda_a,\,\lambda'_b=-\lambda_b,\,\lambda'=-\lambda\,$,
$\eta_a$ and $\eta_b$ are the internal parities of the particles $a$ and $b\,$;
$P$ is the space reflection operator. The amplitudes $F^{(P)}(W,z)$ have the 
same structure as $F(W,z)$ (see Eq.(\ref{h_amp})), but the partial-wave 
amplitudes are taken with the space-reflected initial states. The first factor 
in the right-hand side of Eqs.(\ref{-z}) and (\ref{F-z}) is independent of $j$ 
and, when squared, is always unity, since $(s_a+s_b-\mu)$ is always an integer. 
Of course, these relations could be rewritten in a different form, with space 
reflection affecting the final (instead of initial) states.

Now we can write the differential cross section in two forms:
\be
	\frac{d\sigma(W,-z)}{d z}=\sum_{J=0}^{\infty}\,A^{(\sigma)}_J(W)\,P_J(-z)=
	\sum_{J=0}^{\infty}\,A_J^{(\sigma,\,P)}(W)\,P_J(z)\,,
	\label{crs-z}
\ee
where $A_J^{(\sigma,\,P)}(W)$ has the same structure as $A^{(\sigma)}_J(W)\,$,
but helicity summation uses space-reflected initial (or final) states. Since
$P_J(-z)=(-1)^J P_J(z)\,$, we derive
\be
	A_J^{(\sigma,\,P)}(W)=(-1)^J\,A^{(\sigma)}_J(W)\,.
	\label{coefP}
\ee
If the states used in the summation are separated by their parities, then this
equality means that \mbox{$A^{(\sigma)}_J\,$-coefficients} with odd $J$-values 
may contain only contributions which are bilinear in states of opposite parities. 
For even $J$, on the other hand, bilinear contributions also appear in the 
$A^{(\sigma)}_J$, but only with both states of the same parity, positive or 
negative. Quadratic contributions of any state can appear only at even $J\,$. 
This means, in particular, that proper Breit-Wigner (BW) contributions of a 
resonance of an integer spin $j_{\,R}$ appear in the \mbox{even-$J$} Legendre 
coefficients with $0\leq J\leq2j_{\,R}\,$. But if the spin is half-integer, 
these BW contributions do not appear at $J=2j_{\,R}\,$; they appear only at 
even $J$ with $0\leq J\leq2j_{\,R}-1\,$.

The above expressions clearly demonstrate the well-known statement that the 
unpolarized cross section by itself does not allow a determination of the parity of a 
particular partial wave, since simultaneous reversal of parities for 
\textit{all} partial waves does not influence the cross section. However, if 
there is a resonance with known quantum numbers, including its parity, then 
such complete parity reversal becomes impossible, and even unpolarized cross 
section is able to provide some information on partial-wave parities. Below 
we will discuss this point in more detail.

Described above is the Legendre decomposition for the unpolarized 
differential cross sections. However, such an approach may be applied 
also to processes with polarized particles and/or to polarization 
observables (more exactly, to polarization observables multiplied by the 
unpolarized differential cross section). Such quantities may kinematically 
vanish at $z=\pm1$ (they may even have square root singularities there). 
Decomposition in Legendre polynomials then becomes inadequate, and one 
should instead use Wigner harmonics (or, in particular, associated 
Legendre polynomials) with integer $J$. For example, the beam asymmetry 
studied by the CLAS Collaboration~\cite{CLAS} contains the kinematical 
factor $(1-z^2)$ which automatically arises in any converging series over 
the associated Legendre polynomials $P_J^{\,2}(z)$). In such cases the 
decomposition retains connection (\ref{3j}) between $J$ and $j_1, j_2$; 
relation (\ref{-z}) again allows one to separate interferences of states with 
the same or with opposite parities.

Let us briefly discuss one more point. The series (\ref{h_amp}) and 
(\ref{decomp}) generally sum an infinite number of terms. In practical cases, 
the series will be truncated. This may be justified on the basis of physical 
reasons (\textit{e.g.}, presence of pronounced resonances in the data, 
with known definite spins and parities) or phenomenological ones 
(\textit{e.g.}, higher Legendre coefficient may be safely discarded if their 
fitting errors exceed fitted values). In both cases, we obtain a 
limited number of parameters to describe experimental data and to 
investigate their physical content.

\section{Photoproduction of a Spinless Meson}
\label{sec:photo}

To illustrate the above approach, we consider in more detail the 
particular case of a pseudoscalar-meson photoproduction off the proton, 
for instance,
\be
	\gamma+p \to \pi^+ +n\,.
	\label{pi}
\ee
The initial state has four possible helicity combinations ($\lambda_\gamma
=\pm1,\,\lambda_p=\pm1/2$), while the final state has two helicity 
combinations ($\lambda_\pi=0,\, \lambda_n=\pm1/2$). Thus, there are eight 
different transitions between various initial and final helicities and, 
generally, eight different helicity amplitudes.

It is interesting to emphasize that the value of $\lambda$ unambiguously 
determines all initial helicities: if $\lambda=\pm1/2\,$, then  
$\lambda_\gamma=\pm1\,,\,\lambda_p=\pm1/2\,$; if $\lambda=\pm3/2\,$, then 
$\lambda_\gamma=\pm1\,,\,\lambda_p=\mp1/2$ (of course, this is due to 
absence of $\lambda_\gamma=0\,$). Hence, the independent amplitudes may 
be denoted as $F_{\lambda\pm}(W,z)\,$, where the sign in the index is the 
sign of $\mu\,$, opposite to the sign of helicity of the final nucleon. 
Expression (\ref{unpcrsec}) for the unpolarized cross section may be 
rewritten as
\be
	\frac{d\sigma(W,z)}{d z}=N\sum_{\lambda}\,\left[|F_{\lambda+}(W,z)|^2
	+|F_{\lambda-}(W,z)|^2\right]\,
	\label{phunp}
\ee
with $\lambda$-summation over four values $\pm1/2,\pm3/2\,$. The initial 
state with a particular value of $\lambda$ can be realized by using the 
circularly polarized photon (with a definite helicity) together with the 
longitudinally polarized target nucleon. Therefore, also measurable is the 
cross section for any fixed $\lambda\,$:
\be
	\frac{d\sigma^{(\lambda)}(W,z)}{d z}=4N\left[|F_{\lambda+}(W,z)|^2
	+|F_{\lambda-}(W,z)|^2\right]\,.
	\label{phlam}
\ee
The additional factor 4, as compared to Eq.(\ref{phunp}), arises since
$d\sigma^{(\lambda)}/d z$ deals with a single initial state, while the
unpolarized expression (\ref{phunp}) implies averaging over four initial 
states with different helicities.

Note that amplitudes with all helicities reversed are related by parity
conservation~\cite{jw}, so that only four of the eight amplitudes are 
independent.  Eq.(44) of Ref.~\cite{jw}, applied to the photoproduction reaction, 
gives
\be
	F_{-\lambda\pm}(W,z)=(-1)^{\lambda\mp1/2}\,F_{\lambda\mp}(W,z)\,.
	\label{par}
\ee
Therefore, we can use only amplitudes with positive values of $\lambda=
1/2,\,3/2\,$.  For negative values of $\lambda$ we have
$$
	F_{-1/2\pm}(W,z)=\pm F_{1/2\mp}(W,z)\,,~~~
	F_{-3/2\pm}(W,z)=\mp F_{3/2\mp}(W,z)\,.
$$
As a result,
$$
	\frac{d\sigma^{(-1/2)}(W,z)}{d z}=\frac{d\sigma^{(1/2)}(W,z)}{d z}\,,~~~~
	\frac{d\sigma^{(-3/2)}(W,z)}{d z}=\frac{d\sigma^{(3/2)}(W,z)}{d z}
$$
due to parity conservation. Moreover,
\be
	\frac{d\sigma(W,z)}{d z}=\frac12\left[\frac{d\sigma^{(1/2)}(W,z)}{d z}
	+\frac{d\sigma^{(3/2)}(W,z)}{d z}\right]\,.
	\label{sig-obs}
\ee
Difference of the helicity cross sections is related to the double polarization
observable $E\,$:
\be
	\check{E}(W,z)=E\cdot\frac{d\sigma(W,z)}{d z}=
	\frac12\left[\frac{d\sigma^{(1/2)}(W,z)}{d z}
	-\frac{d\sigma^{(3/2)}(W,z)}{d z}\right]\,.
	\label{e-obs}
\ee
Following Walker~\cite{Walker}, if we let $H_1$ to $H_4$ label the
four independent helicity amplitudes,   
the translation to amplitudes of the form
$A_{\mu\lambda}(W,z,)$ is given in Table~\ref{tab:tab1}.
\begin{table}[htb!]
\begin{center}
\caption{\label{tab:tab1} Walker notation~\cite{Walker} for 
helicity amplitudes $A_{\mu\lambda}(W,z)$.}
\begin{tabular}{|c|cc|cc|}
\hline
$\lambda\rightarrow$  & $\lambda_{\gamma}=+1$  &               & $\lambda_{\gamma}=-1$ & \\
$\mu\downarrow$       & $\frac{3}{2}$   & $\frac{1}{2}$ & $-\frac{1}{2}$ & $-\frac{3}{2}$ \\
\hline
 $\frac{1}{2}$& H$_1$    & H$_2$    & H$_4$     &-H$_3$ \\
$-\frac{1}{2}$& H$_3$    & H$_4$    &-H$_2$     & H$_1$ \\
\hline
\end{tabular}
\end{center}
\end{table}

Partial-wave decomposition of the amplitudes $F_{\lambda\pm}(W,z)$ contains 
the partial-wave helicity amplitudes which may be analogously denoted as 
$f^j_{\lambda\pm}(W)\,$, with the same meaning of indices; note that $j\geq1/2$ 
for $f^j_{1/2\pm}(W)$, while $j\geq3/2$ for $f^j_{3/2\pm}(W)\,$.

Further, the helicity partial-wave amplitudes can be combined so to obtain two 
sets of definite-parity partial-wave amplitudes $f_{\lambda}^{j\pm}(W)$. 
According to Eq.(41) of Ref.~\cite{jw}, we obtain
\be
	\sqrt{2}\,f_{\lambda}^{j\pm}(W)=f^j_{\lambda+}(W)\pm\,\eta_\pi 
	\eta_N\,(-1)^{j-1/2} f^j_{\lambda-}(W)\,,
\ee
where $\lambda=1/2$ or 3/2\,, the upper sign $\pm$ corresponds to the final 
(and initial as well) state parity equal to $\pm1\,$, $\eta_\pi$ and $\eta_N$ 
are intrinsic parities of the pion and nucleon. The inverted expressions are
\be
	\sqrt{2}\,f_{\lambda+}^j(W)=f^{j+}_{\lambda}(E)+f^{j-}_{\lambda}(W)\,,~~~
	\sqrt{2}\,\eta_\pi \eta_N\,(-1)^{j-1/2}f^j_{\lambda-}(E)=f^{j+}_{\lambda}(W)-
	f^{j-}_{\lambda}(W)\,.
\ee
Of course, $\eta_\pi \eta_N=-1,\,(\eta_\pi \eta_N)^2=1\,$. Recall also that
the lower sign $+$ or $-$ corresponds to the final state value $\mu=\pm1\,$,
while the upper sign corresponds to parity of the final (and initial) state.

It is easy to check that
\be
	f_{\lambda_1+}^{j_1*}\,f_{\lambda_2+}^{j_2}\pm
	(-1)^{j_1+j_2+1}f_{\lambda_1-}^{j_1*}\,f_{\lambda_2-}^{j_2}=
	f_{\lambda_1}^{j_1+*}\,f_{\lambda_2}^{j_2\pm}+
	f_{\lambda_1}^{j_1-*}\,f_{\lambda_2}^{j_2\mp}\,,
\ee
\be
	(-1)^{j_2+1/2}\,f_{\lambda_1+}^{j_1*}\,f_{\lambda_2-}^{j_2}\pm
	(-1)^{j_1+1/2}\,f_{\lambda_1-}^{j_1*}\,f_{\lambda_2+}^{j_2}=
	f_{\lambda_1}^{j_1\pm*}\,f_{\lambda_2}^{j_2+}-
	f_{\lambda_1}^{j_1\mp*}\,f_{\lambda_2}^{j_2-}\,.
\ee
In particular,
$$
	|f^j_{\lambda+}|^2+|f^j_{\lambda-}|^2=|f_{\lambda}^{j+}|^2+
	|f_{\lambda}^{j-}|^2\,,~~~
	|f^j_{\lambda+}|^2-|f^j_{\lambda-}|^2=2\,\mathrm{Re}(f_{\lambda}^{j+*}\,
	f_{\lambda}^{j-})
$$
(recall that here all the $j$-values are half-integer, so $j_1+j_2+1$ is always 
an integer).
The translation from these partial-wave amplitudes to the helicity elements, 
$A_{\ell \pm}$ and $B_{\ell \pm}$,
as well as the multipole amplitudes, $E_{\ell \pm}$ and $M_{\ell \pm}$, 
is given in Ref.~\cite{Walker}. For
example, we have
\begin{align}
-f_{1/2}^{1/2-} &=  A_{0+} = E_{0+}~, \\
f_{1/2}^{1/2+} &=   A_{1-} = M_{1-}~,
\end{align}
where the subscript notation $\ell \pm$ for helicity elements and 
multipoles~\cite{CGLN} denotes a state with orbital 
angular momentum $\ell$ and total angular momentum $j=\ell \pm 1/2$. 

The above analysis is equally applicable for the process of the $\eta$-meson 
photoproduction
\be
	\gamma+p \to \eta+p\,
	\label{eta}
\ee
(or for $\pi^0$ photoproduction). The energy region of $\eta$ 
production, investigated experimentally in Ref.~\cite{CBC}, is assumed to contain $N^\ast$ 
resonances with spins up to 5/2~\cite{PDG}. One can expect, therefore, that 
the decompositions should essentially run up to $j=5/2$ in Eq.(\ref{h_amp}) 
and up to $J=5$ in Eq.(\ref{decomp}). Such an 
expectation agrees with the fit to 
data~\cite{CBC}: extracted Legendre coefficients with $J>5$ appear to be 
consistent with zero, within their uncertainties. The Wigner harmonics 
necessary for the amplitude decompositions are given explicitly in 
Appendix~1C.  (Note that for the $\pi^0$ production, at 
similar energies~\cite{A2,CLAS}, one needs more lengthy decompositions up 
to $J=10\,$, because of the lower associated threshold.)

Now we can illustrate our approach in more detail for the cross sections of 
the reaction (\ref{eta}). We use Eq.(\ref{unpcrsec}) (truncated up to 
$j_{1,2}= 5/2\,$) and decompositions of Appendix~1D to derive 
expressions for the Legendre coefficients $A^{(\sigma)}_J~~(J=0,...,5)$. 
They are shown in Appendix~2A. Note that $A^{(\sigma)}_0$ can be 
rewritten in the form
\be
	\frac1{2N}\,A^{(\sigma)}_0(W)=\sum_{j\geq 1/2}(2j+1)\,\left(\,|f_{1/2}^{j+}|^2+
	|f_{1/2}^{j-}|^2\right)+\sum_{j\geq 3/2}(2j+1)\,\left(\,|f_{3/2}^{j+}|^2+
	|f_{3/2}^{j-}|^2\right)\,;
	\label{a0}
\ee
the left-hand side factor 1/2 accounts for the fact that the right-hand side 
expression contains contributions of only one sign of the total helicities, 
$+1/2$ and $+3/2$, while $A^{(\sigma)}_0$ should contain also contributions 
with the negative sign of helicities (recall that positive and negative sign 
contributions are equal to each other, due to parity conservation). This 
relation is true even without \mbox{$j$-truncating} and clearly shows that 
$A^{(\sigma)}_0$ is indeed proportional to the total (\textit{i.e.}, 
integrated differential) cross section, as it should be.

Explicit expressions of Appendix~2A confirm the properties of the Legendre 
coefficients formulated above. Every coefficient has two parts, corresponding 
to $\lambda=1/2$ and $\lambda=3/2$. Of course, states with $j=1/2$ contribute 
only to $\lambda=1/2$. Coefficients $A^{(\sigma)}_J$ with even $J$ consist of 
proper contributions of various partial-wave states and interference 
contributions of states with different $j$, but with the same parities, both 
positive and negative. On the other hand, the odd-$J$ coefficients contain 
only interferences between states of the same or different values of $j$, but 
always with opposite parities, exactly as stated in the preceding Section. 
As the value of $J$ increases, so does the number of contributions to 
$A^{(\sigma)}_J$. The simple structure of $A^{(\sigma)}_4$ and 
$A^{(\sigma)}_5$ in Appendix~2 is the result of our assuming the absence of 
states with $j>5/2$. In particular, it is because of this assumption that 
states with $j=1/2$ are not seen in the displayed expressions for $A^{(\sigma)}_4$ 
and $A^{(\sigma)}_5$.

This approach can be easily applied to other polarization
observables. For example,
the double polarization observable $\check{E}(W,z)=E\cdot d\sigma/dz$ may also 
be expanded in Legendre polynomials $P_J(z)$, similar to Eq.(\ref{decomp}), 
but with different Legendre coefficients $A^{(E)}_J(W)$. Comparison of 
expressions (\ref{sig-obs}) and (\ref{e-obs}) shows that $A^{(\sigma)}_J(W)$ 
and $A^{(E)}_J(W)$ differ only in the sign of all contributions with 
$\lambda=3/2\,$.
The beam-polarizaton quantity, $\Sigma$, is also treated within the helicity
formalism in Ref.~\cite{Walker}, 
\be
	\Sigma(W,z)\cdot\frac{d\sigma(W,z)}{dz}=
	4N\,\textrm{Re}\left[F_{1/2-}^\ast(W,z)\,F_{3/2+}(W,z)-
	F_{1/2+}^\ast(W,z)\,F_{3/2-}(W,z)\right]\,.
	\label{sigsec}
\ee
Both terms in the square brackets contain the kinematical edge factor 
$(1-z^2)$, and we expand them over the associated Legendre functions $P^2_J(z)$ 
with $J\geq2$, which all have the same edge factor:
\be
	\hat{\Sigma}(W,z)\equiv\Sigma(W,z)\cdot\frac{d\sigma(W,z)}{dz}=
	\sum_{J=2}^{\infty}\,A^{(\Sigma)}_J(W)\,P^2_J(z)\,.
	\label{sigsecJ}
\ee
Application of property (\ref{F-z}) to the right-hand side of expression 
(\ref{sigsec}) shows that the coefficients $A^{(\Sigma)}_J(W)$ also contain 
parity correlations, just in the same way as the coefficients 
$A^{(\sigma)}_J(W)$ for the unpolarized cross section.

Expansions for products of $d$-harmonics, up to $j=5/2$, arising in 
Eq.(\ref{sigsec}) are given in Appendix~1E. They allow to obtain 
$A^{(\Sigma)}_J(W)$ with $J\leq5$ truncated at $j=5/2$.  The results are 
shown in Appendix~2B. All contributions to the Legendre coefficients 
$A^{(\Sigma)}_J(W)$ come from interferences of amplitudes with initial 
total helicities 1/2 and 3/2. State parities are correlated exactly as 
for $A^{(\sigma)}_J(W)$: the same parities in the even-$J$ coefficients 
and opposite parities in the odd-$J$ ones. There is, however, an 
interesting difference, not quite evident in expression (\ref{sigsec}). 
Inversion of parities for all states does not change $A^{(\sigma)}_J(W)$ 
and, therefore, the cross sections (differential and total). On the other 
hand, such a transformation reverses the signs of all $A^{(\Sigma)}_J(W)$ and, 
therefore, of $\hat{\Sigma}(W,z)$ and $\Sigma(W,z)$ as well.

\section{Application to Data}
\label{sec:app}

The expansion method requires data of both high precision and broad
angular coverage to determine the higher-order coefficients. 
A prime example is provided by 
the A2 Collaboration at MAMI which recently reported
7978 $d\sigma/d\Omega$ data for the reaction $\gamma p\to\pi^0p$
and for for incident photon energies $E$ from 218 to 1573~MeV
(or for c.m energies W = 1136 -- 1957~MeV)~\cite{A2}. These data are
obtained with a fine binning in $E$ ($\sim$4~MeV for all energies
below E = 1120~MeV) and 30 angular bins, giving a good coverage of 
the $\pi^0$ production angle.  The data obtained above E = 1443~MeV
(W = 1894~MeV), however, have a limited angular coverage and for this
reason were excluded from the present Legendre fit.

A good description of the $\gamma p\to
\pi^0p$ differential cross sections was obtained, for each included energy bin
and the full angular range, in a fit with Legendre
polynomials up to order ten (Eq.(\ref{decomp})) - the
coefficients $A^{(\sigma)}_J$ depending on energy.

\begin{figure*}[htb!]
\begin{center}
	\includegraphics[height=5.5cm, keepaspectratio, angle=90]{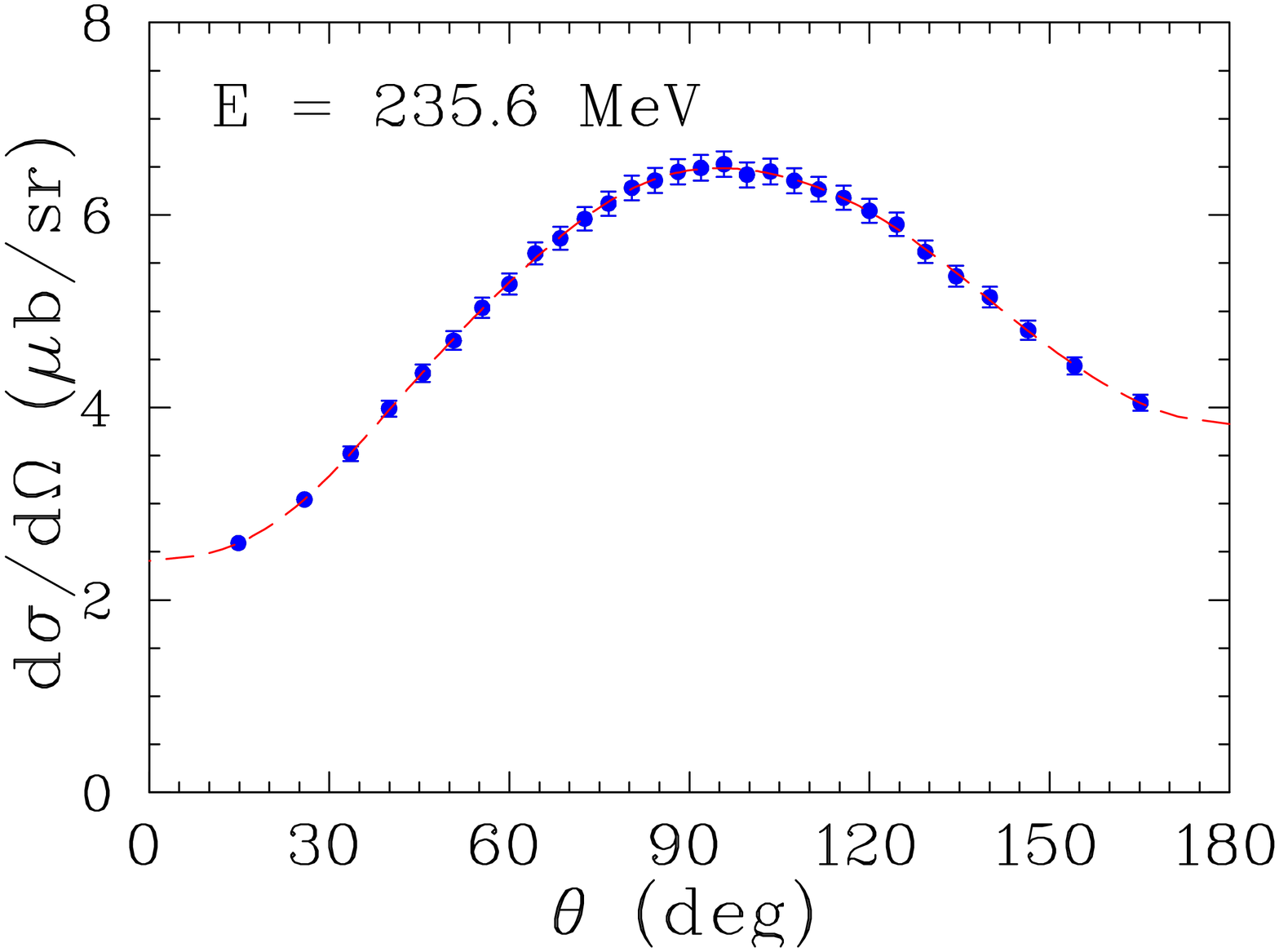}
        \includegraphics[height=5.5cm, keepaspectratio, angle=90]{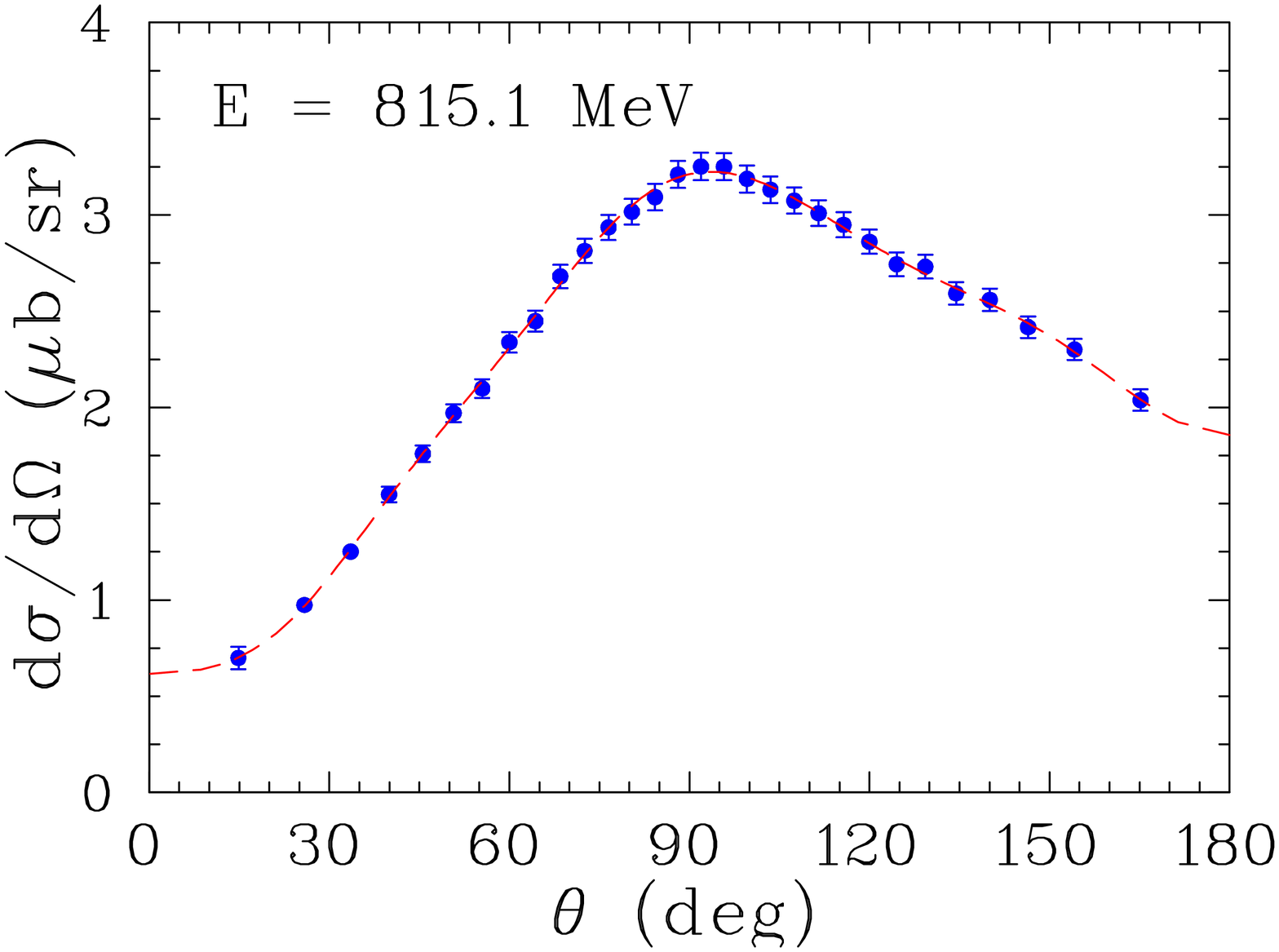}
        \includegraphics[height=5.5cm, keepaspectratio, angle=90]{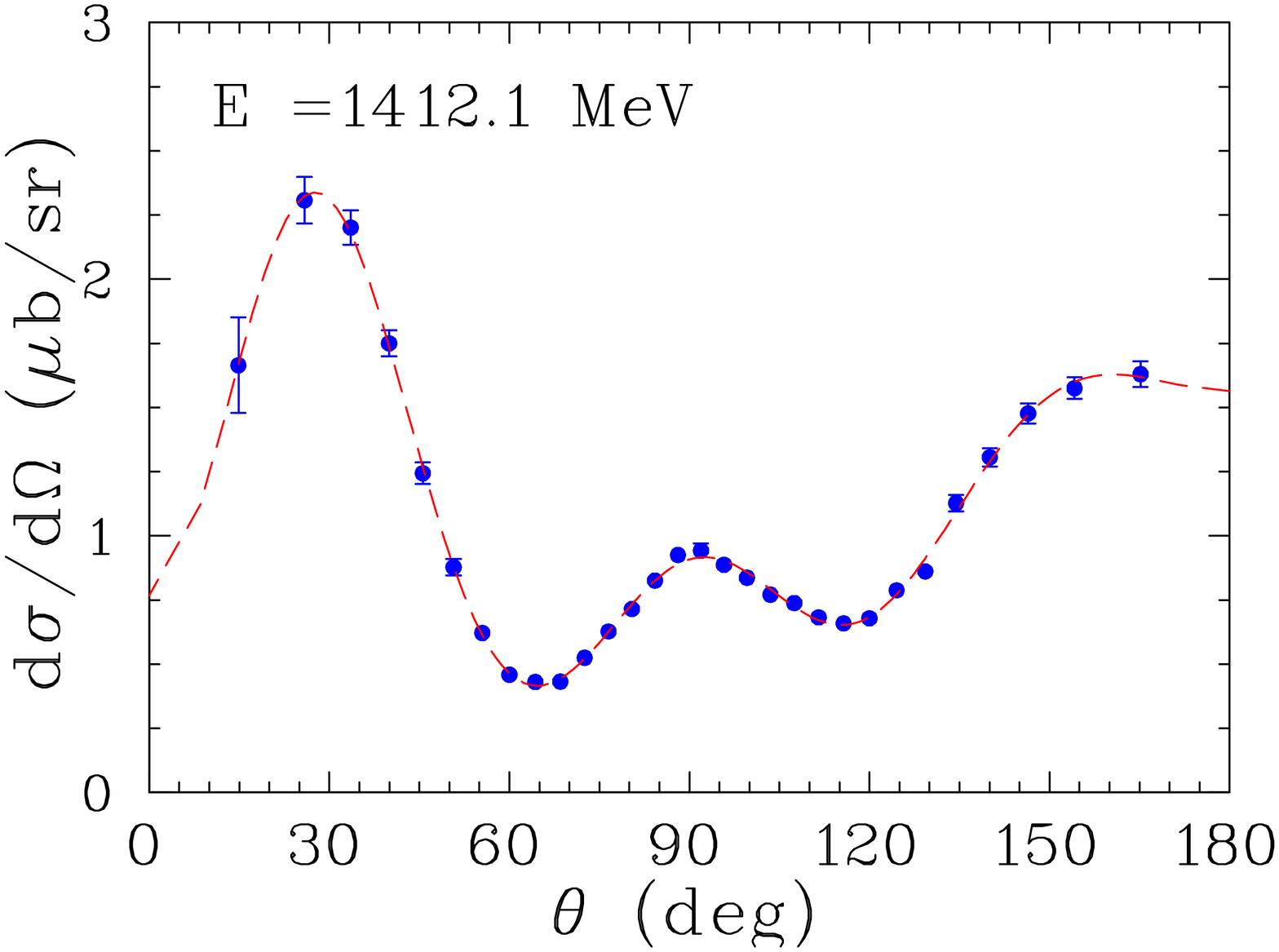}
\end{center}
\begin{center}
        \includegraphics[height=6in, keepaspectratio, angle=90]{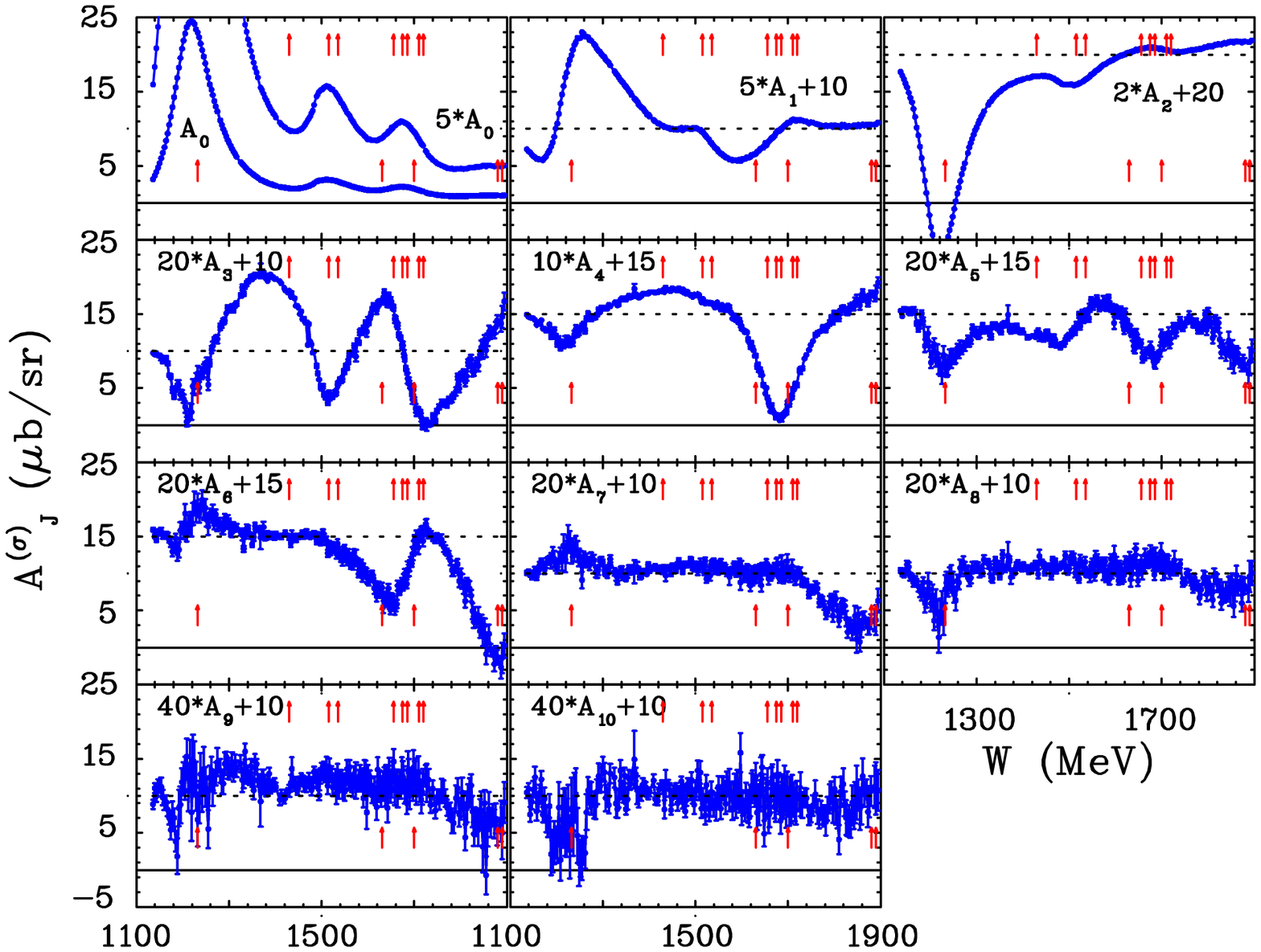}
\end{center}

        \caption {(Color online) 
		\textit{Top panel}: Samples of the $\gamma p\to\pi^0p$ 
		differential cross sections, $d\sigma/d\Omega$, from 
		A2 Collaboration 
		at MAMI measurements (blue filled circles)~\protect\cite{A2} 
		with the best fit results using Legendre polynomials 
		(red dashed lines). The 
		error bars on all data points represent statistical 
		uncertaities only. Values of $E$ in each plot indicate 
		the lab photon energies. 
		\textit{Bottom panel}: Coefficients of Legendre 
		polynomials (blue filled circles). The error bars of 
		all values represent A$^{(\sigma)}_J$ uncertainties 
		from the fits in which only the statistical uncertainties 
		were used. Solid lines are plotted to help guide 
		the eye. Red vertical arrows indicate masses of the 
		four-star resonances (BW masses) known in this energy 
		range~\protect\cite{PDG}. The upper row of arrows 
		corresponds to N$^\ast$ states with isospin $I = 1/2$ and the 
		lower row corresponds to $\Delta^\ast$ with $I = 3/2$.}
		\label{fig:legA2}
\end{figure*}


The typical experimental statistics and the Legendre-polynomial 
fits are illustrated in Fig.~\ref{fig:legA2}(top panel) for 
different energies.  The results of the Legendre-polynomial fits 
for each coefficient $A^{(\sigma)}_J$ are depicted in 
Fig.~\ref{fig:legA2}(bottom panel), showing their energy 
dependence in unprecedented detail. 

As expected from the form of Eq.(\ref{a0}), resonance contributions
from the first, second and third resonance regions combine to
produce clear peaks in the coefficient $A^{(\sigma)}_0$.
These regions are somewhat less pronounced in $A^{(\sigma)}_2$, which
also contains interference terms between states of the same parity.
The result for $A^{(\sigma)}_0$ itself shows 
good agreement with the total cross section, 
obtained by an integration of the differential cross sections, 
confirming the quality of this dataset.

\begin{figure*}[htb!]
        \includegraphics[height=5in, keepaspectratio, angle=90]{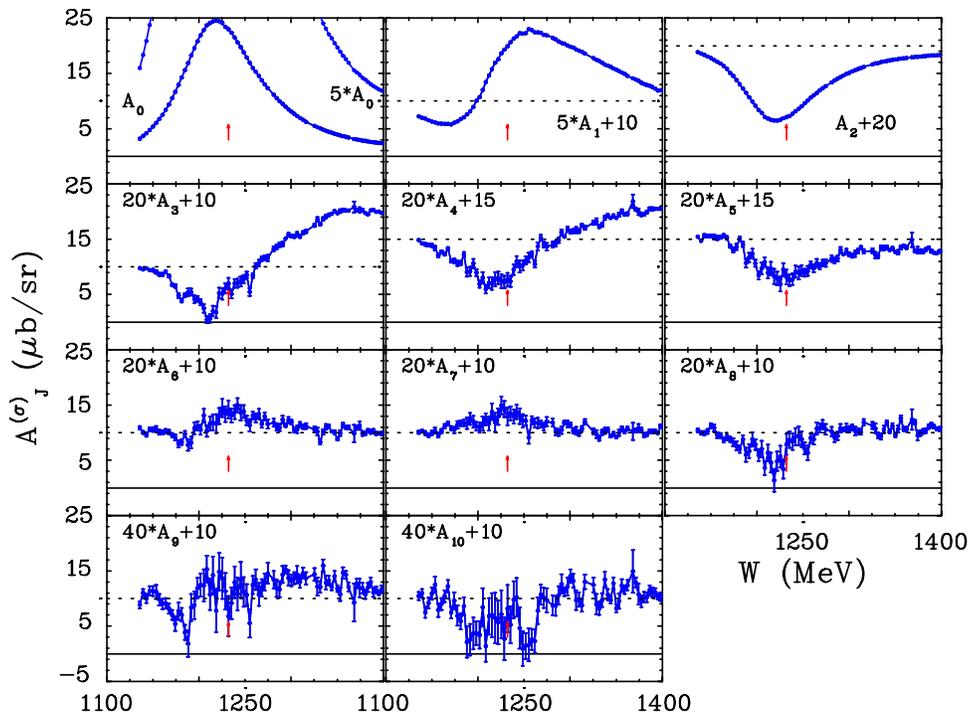}

        \caption {(Color online) Zoom for A2 $\gamma p\to\pi^0p$
                data below W = 1400~MeV to cover the $\Delta$-isobar
                region~\protect\cite{A2} as shown on
                Fig.~\protect\ref{fig:legA2}(bottom panel).}
                \label{fig:legA2b}
\end{figure*}

Other interesting features in Fig.~\ref{fig:legA2}, and the expanded
plot of Fig.~\ref{fig:legA2b}, are the sharp structures seen for each 
coefficient $A^{(\sigma)}_J$ in the region of $\Delta(1232)3/2^+$.  
This resonance can contribute directly (without interference) only to 
$A^{(\sigma)}_0$ and $A^{(\sigma)}_2$. Since there exists no other 
nearby resonance, such structures can appear only due to the 
interference of $\Delta(1232)3/2^+$ with other non-resonant 
partial-waves. Coefficient $A^{(\sigma)}_1$ should reveal the 
interference of $\Delta(1232)3/2^+$ with the three states having 
$J^P = 1/2^-$, $3/2^-$, and $5/2^-$. Higher coefficients 
$A^{(\sigma)}_J$ could reveal interference with the four states 
having $J = j - 3/2, j - 1/2, j + 1/2$, and $j + 3/2$, with parity 
positive for even $j$ and negative for odd $j$. Thus, via such 
interference effects, the contributions from very high partial-wave 
amplitudes could be studied, a possibility not available in any 
other standard approach. This feature is similar to enhancing the
manifestation of rare decay modes of resonances via the interference 
with other strong resonances~\cite{YaA}.

It should be noted that the recurring sharp structures associated with
the $\Delta (1232)$ energy region do not appear in multipole analyses
of these data. The angle-independent systematic error was used to determine
renormalization factors for each angular distribution. These factors were
determined to be very near unity (within 1\%) and, if applied to the
data, had no effect on the higher Legendre coefficients. If instead,
the statistical and systematic errors are added in quadrature~\cite{Beck},
structures in the highest coefficients are masked by greatly expanded errors. 
This result emphasizes the importance of systematic error analysis, the 
effect of which may also be magnified by a dominant resonance.

\begin{figure*}[htb!]
\begin{center}
        \includegraphics[height=5.5cm, keepaspectratio, angle=90]{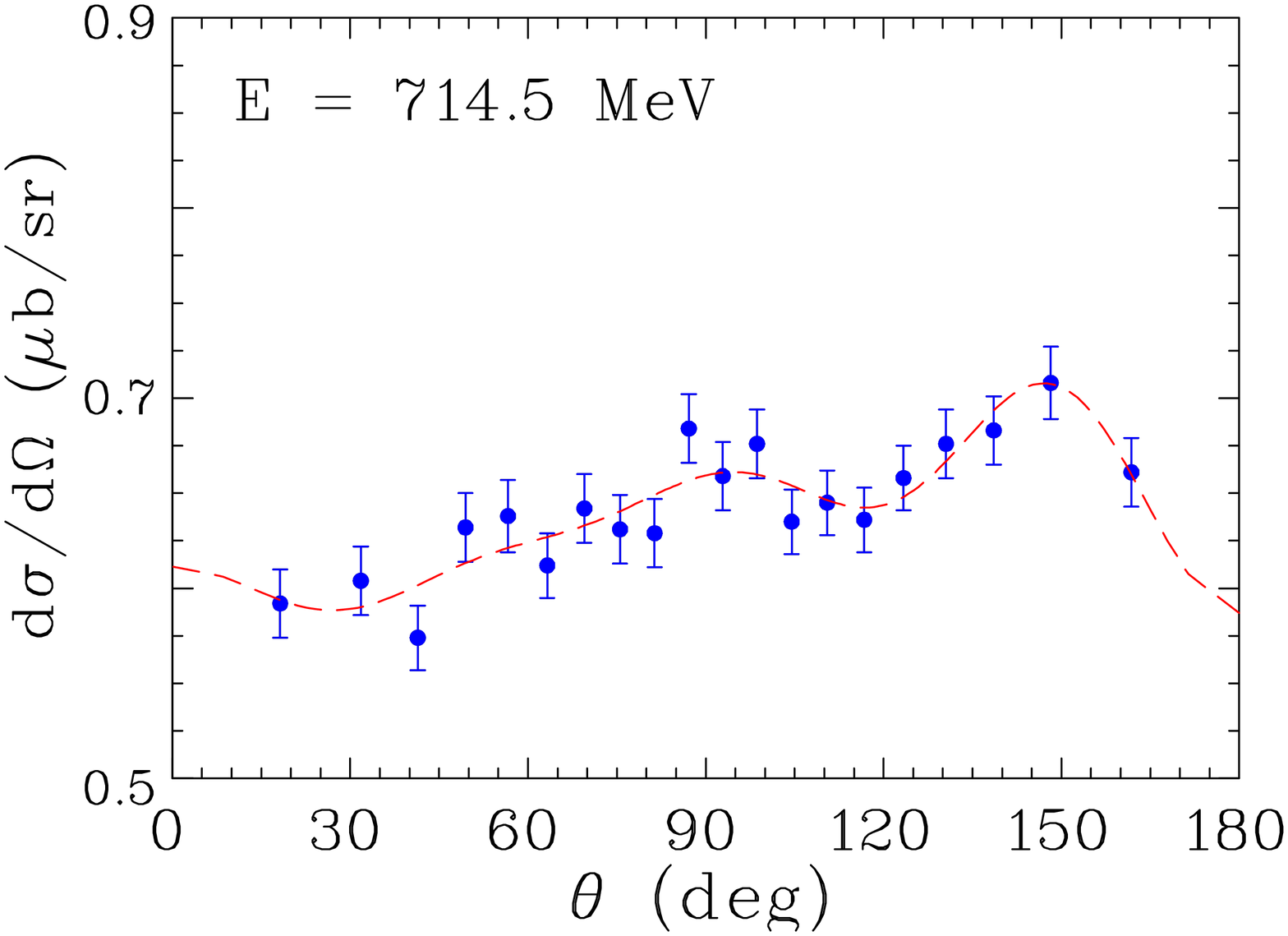}
        \includegraphics[height=5.5cm, keepaspectratio, angle=90]{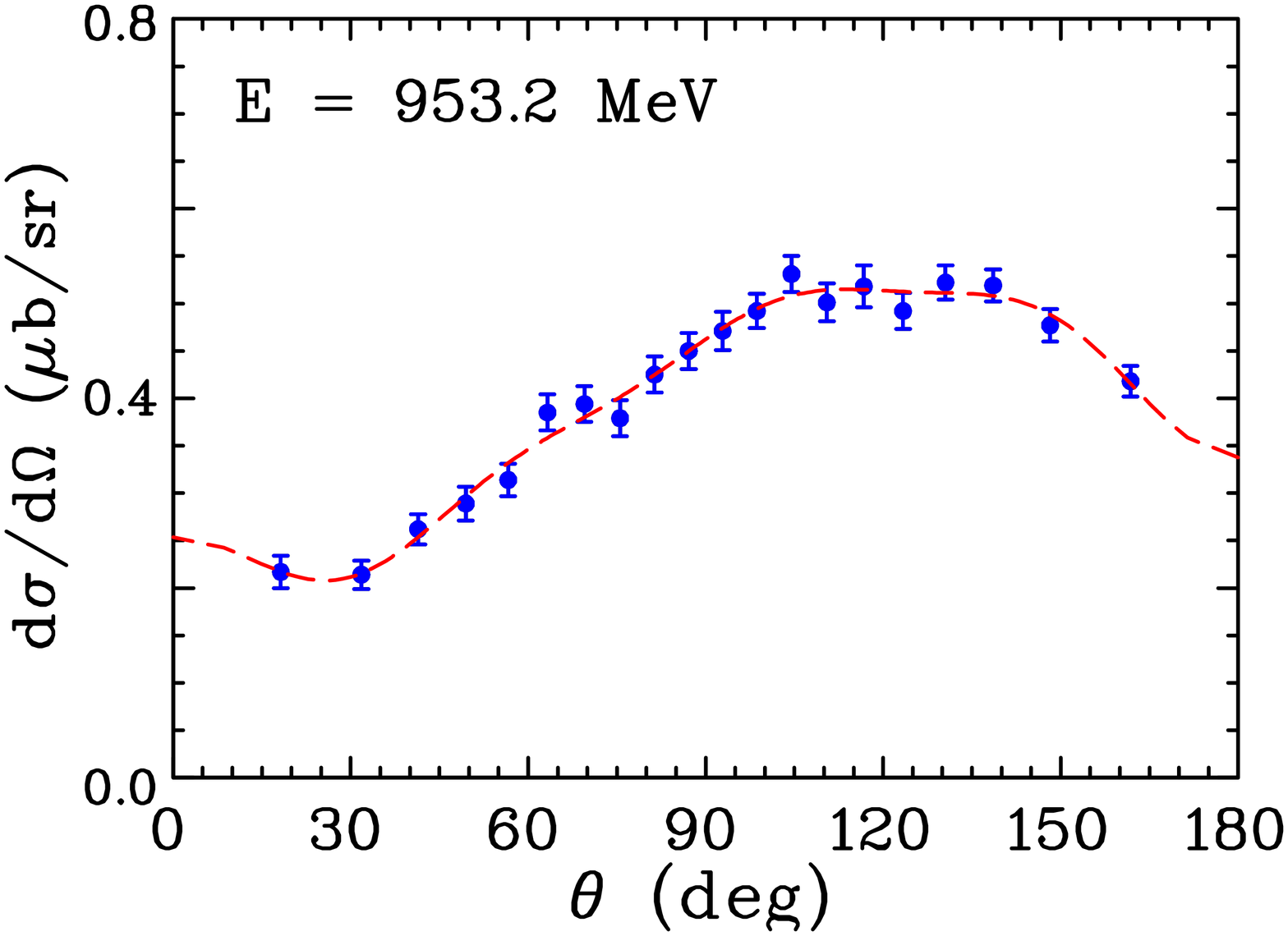}
        \includegraphics[height=5.5cm, keepaspectratio, angle=90]{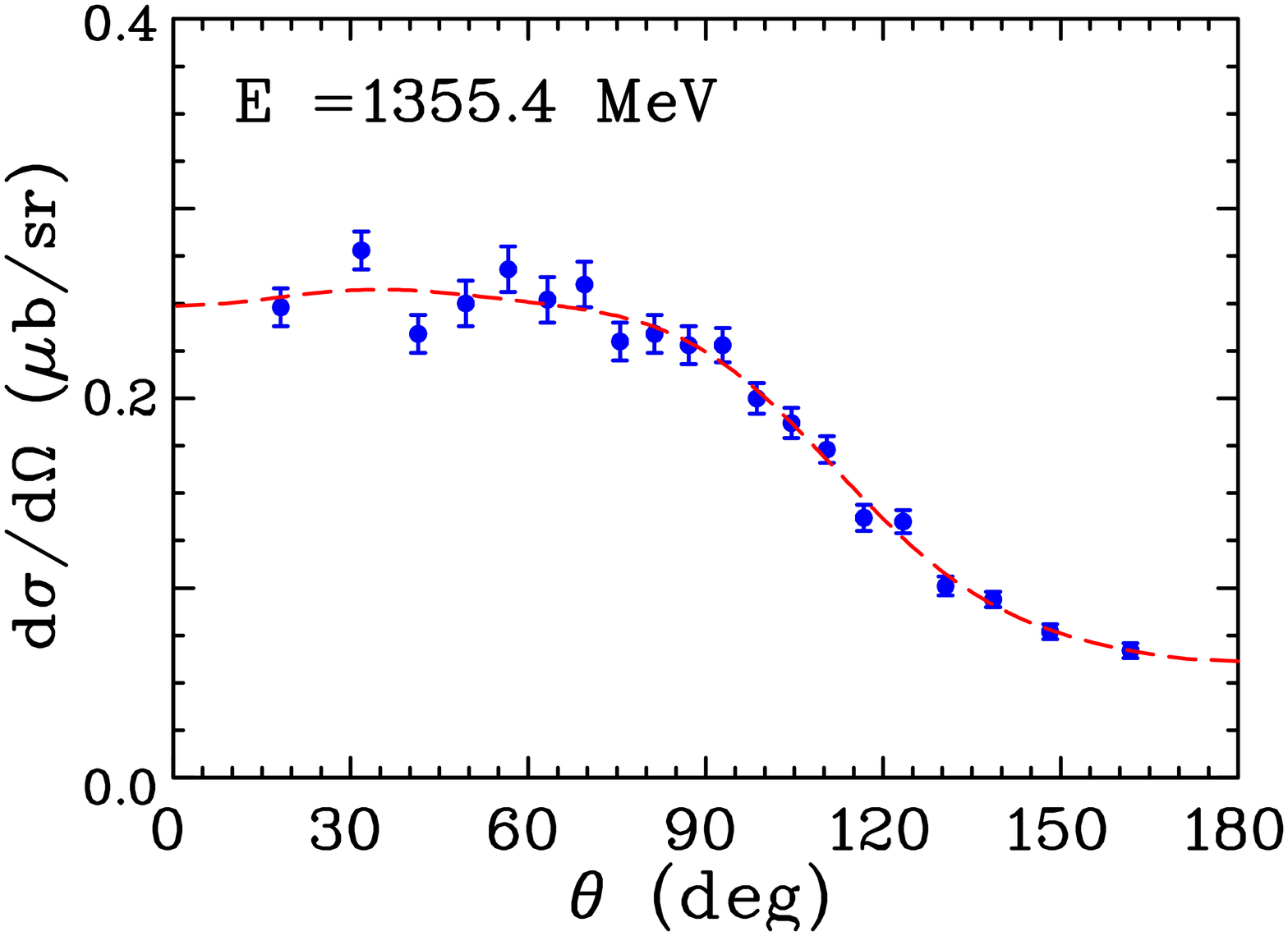}
\end{center}
\begin{center}
        \includegraphics[height=6in, keepaspectratio, angle=90]{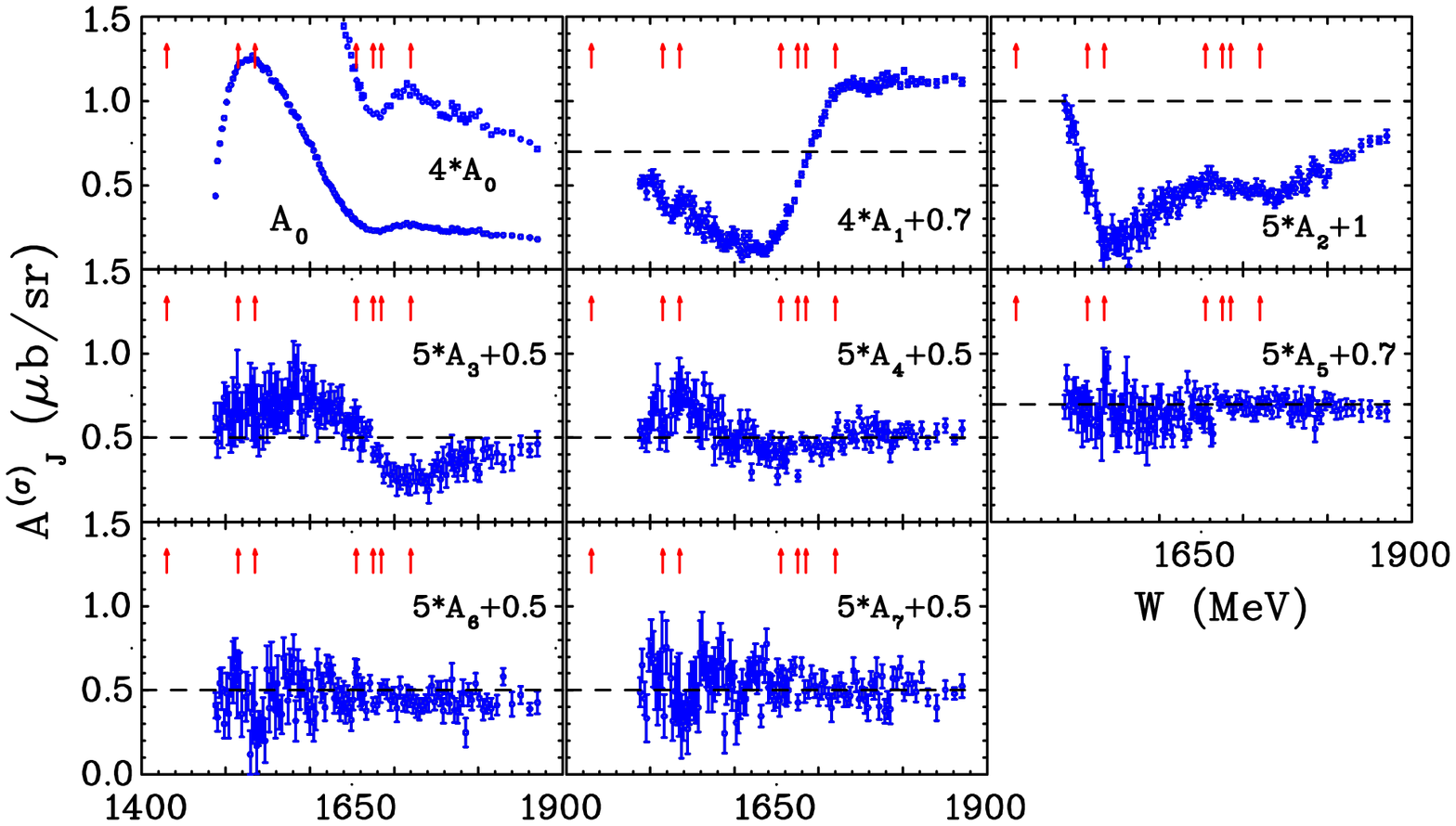}
\end{center}

        \caption {(Color online)
                \textit{Top panel}: Samples of the $\gamma p\to\eta p$
                differential cross sections, $d\sigma/d\Omega$, from 
		A2 Collaboration
                at MAMI measurements~\protect\cite{CBC} with the
                best fit results using Legendre polynomials. The 
		notation as given
                in Fig.~\protect\ref{fig:legA2}.
                \textit{Bottom panel}: Coefficients for Legendre
                polynomials.  The notation is given in
                Fig.~\protect\ref{fig:legA2}.} \label{fig:legA2a}
\end{figure*}

The A2 collaboration at MAMI has also measured
2400 $d\sigma/d\Omega$s for the reaction $\gamma p\to\eta p$ and
for for incident photon energies $E$ from 710 and up to 1395~MeV
(or for c.m energies W = 1488 -- 1870~MeV]~\cite{CBC}. The large
number of events accumulated allowed the division of the data
into 120 bins in $E$. From the reaction threshold to an $E$ of
1008~MeV, the bin width was that of a single tagger channel
($\sim$4~MeV). From 1008 to 1238~MeV, two tagger channels were
combined to a single energy bin.  Above 1238~MeV, an energy bin
included from three to eight tagger channels. The $\gamma p\to\eta 
p$ differential cross sections were determined as a function of
$z$. The $z$ spectra at all energies were divided into 20 bins.

The photoproduction of eta mesons is interesting in that only
isospin 1/2 resonances can contribute, thus reducing the list
of candidates required to explain energy-dependent structures in
the Legendre coefficients. For many $N^*$ states, the decay to
$\eta N$ has been determined to be very weak. This too helps
in deciphering the sources of structures. 

In Fig.~\ref{fig:legA2a}(top panel), differential cross sections
for three incident photon energies are compared with the
Legendre-polynomial fits.  The results of the Legendre-polynomial
fits for each coefficient $A^{(\sigma)}_J$ are depicted in
Fig.~\ref{fig:legA2a}(bottom panel), showing their energy
dependence.  The full angular coverage of A2 differential cross
sections together with small statistical uncertainties allowed
a reliable determination of several Legendre coefficients
$A^{(\sigma)}_J$, which was difficult to achieve with the previous
data. 

The behavior of these Legendre coefficients suggests possible
resonance contributions, though some puzzles remain. Unlike the
pion photoproduction case, $A^{(\sigma)}_0$ reveals only one dominant
resonance (N(1535)$1/2^-$) with a small shoulder near 1700 MeV,
possibly containing several states. While the $\Delta (1232)$ state
appeared prominently also in $A^{(\sigma)}_2$, the lower-spin 
N(1535) does not. Instead, near threshold there is a nearly linear
drop from zero, which must involve an interference with the dominant
N(1535). From Appendix 2, likely states have $J^P$ = $3/2^-$  
and $5/2^-$.

Perhaps the most intriguing structure is seen in $A^{(\sigma)}_1$. 
Assuming this is due to states, with opposite parity, interfering
with the tail of the dominant N(1535), candidates include
$J^P$ = $1/2^+$ and $3/2^+$. The crossover seen, less
clearly, in the coefficient $A^{(\sigma)}_3$ nearly mirrors that
found in $A^{(\sigma)}_1$, suggesting a common origin. 

\section{Discussion and Conclusions}
\label{sec:conc}

Several examples of the Legendre analyses, discussed in this paper, 
are rather simple.  They, nevertheless, allow us to demonstrate various 
features, inherent also in more general and complicated cases. That is 
why we are now able to formulate a number of sufficiently general 
conclusions.

\begin{itemize}
\item Legendre expansions provide a model-independent approach suitable for 
	presentation of modern detailed (high-precision and high-statistics) 
	data for two-hadron reactions.
\item This approach is applicable both to cross sections and to polarization 
	variables; it is much more compact than traditional methods, at 
	least, at energies within the resonance region.
\item The Legendre coefficients reveal specific correlations and 
	interferences between states of definite parities.
\item Due to interference with resonances, high-momentum Legendre 
	coefficients open a window to study higher partial-wave amplitudes, 
	which are out-of-reach within any other method.
\end{itemize}

Concluding this brief discussion, one should emphasize that direct 
interference has become a useful instrument to search for and study 
rare decays of well-established resonances. However, its possibilities are 
limited by restrictions for the resonance quantum numbers. Rescattering 
interference is not limited by such requirements and, therefore, may 
provide effective methods to search for and study new resonances with arbitrary 
quantum numbers. Data on multi-hadron decays of heavy particles also 
present a new rapidly-expanding area for applications of different kinds 
of interference both to study spectroscopy of resonances and to establish 
their characteristics.

\vspace{5mm}
\centerline{\bf Acknowledgments}
\vspace{2mm}

Ya.~I.~A. acknowledged support by the Russian Science Foundation
(Grant No.~14--22--00281); the work of W.~J.~B. and I.~I.~S. is supported,
in part, by the U.~S. Department of Energy, Office of Science, Office
of Nuclear Physics, under Awards No.~DE--SC0014133 and DE--SC0016583.
R.~L.~W. is supported by the U.~S. Department of Energy Grant DE--SC0016582.

\vspace{5mm}
\centerline{\bf Appendix~1.~Wigner harmonics and Legendre functions}
\vspace{2mm}

For convenience, here we give explicitly those Legendre functions
and Wigner harmonics which can be used to describe differential cross
sections and beam asymmetries for reactions (\ref{pi}) and (\ref{eta})
up to $j, J=5$.

A) Legendre functions $P_J(z)$ with $J\leq5\,$:
$$ 
	P_0(z)=1,~~~P_1(z)=z,~~~P_2(z)=\frac{1}{2}(3z^2-1),~~~
	P_3(z)=\frac{1}{2}(5z^3-3z),   
$$
$$ 
	P_4(z)=\frac{1}{8}(35z^4-30z^2+3),~~~
	P_5(z)=\frac{1}{8}(63z^5-70z^3+15z)\,.  
$$

B) Associated Legendre functions $P^2_J(z)$ with $J\leq5\,$:
$$ 
	\frac{P^2_2(z)}{(1-z^2)}=3,~~~\frac{P^2_3(z)}{(1-z^2)}=15z,~~~
	\frac{P^2_4(z)}{(1-z^2)}=\frac{15}2(7z^2-1),~~~
	\frac{P^2_5(z)}{(1-z^2)}=\frac{105z}2(3z^2-1)\,.
$$

C) Wigner harmonics $d^{\,j}_{\lambda,\mu}(z)$ with $j\leq5/2$ for
$\lambda=\pm3/2, \pm1/2$ and $\mu=\pm1/2\,$.\\
$$ 
	~d^{\,j}_{-1/2,-1/2}(z)=d^{\,j}_{1/2,1/2}(z)\,,~~~~~
	d^{\,j}_{-1/2,1/2}(z)=-d^{\,j}_{1/2,-1/2}(z)=(-1)^{j-1/2}\,
	d^{\,j}_{1/2,1/2}(-z)\,,
$$
$$ 
	d^{\,j}_{-3/2,-1/2}(z)=-d^{\,j}_{3/2,1/2}(z)\,,~~~
	d^{\,j}_{-3/2,1/2}(z)=d^{\,j}_{3/2,-1/2}(z)=(-1)^{j-1/2}\,
	d^{\,j}_{3/2,1/2}(-z)\,.
$$
It is sufficient, therefore, to know explicitly only $d$-functions
with both $\lambda$ and $\mu$ positive.
a) The case of $\lambda=1/2\,$:
$$ 
	d^{\,1/2}_{1/2,1/2}(z)=\sqrt{\frac{1+z}2}\,,~~~
	d^{\,3/2}_{1/2,1/2}(z)=\frac12\,\sqrt{\frac{1+z}2}\,(3z-1)\,,~~~
	d^{\,5/2}_{1/2,1/2}(z)=\frac12\,\sqrt{\frac{1+z}2}\,(5z^2-2z-1)\,.
$$
b) The case of $\lambda=3/2\,$:
$$ 
	d^{\,3/2}_{3/2,1/2}(z)=-\frac{\sqrt{3}}2\,\sqrt{\frac{1-z}2}\,(1+z)\,,~~~
	d^{\,5/2}_{3/2,1/2}(z)=-\frac{\sqrt{2}}4\,\sqrt{\frac{1-z}2}\,(1+z)(5z-1)
\,.
$$

D) Expansions (for cross sections) over Legendre functions.\\
a) Quadratic terms with $\lambda=1/2\,$:
$$ 
	[\,d^{\,1/2}_{1/2,1/2}(z)\,]^2=\frac12\,P_1(z)+\frac12\,P_0(z)\,,
$$
$$ 
	[\,d^{\,3/2}_{1/2,1/2}(z)\,]^2=\frac9{20}\,P_3(z)+\frac14\,P_2(z)+
	\frac1{20}\,P_1(z)+\frac14\,P_0(z)\,,
$$
$$ 
	[\,d^{\,5/2}_{1/2,1/2}(z)\,]^2=\frac{25}{63}\,P_5(z)+\frac17\,P_4(z)+
	\frac4{45}\,P_3(z)+\frac4{21}\,P_2(z)+\frac1{70}\,P_1(z)+\frac16\,P_0(z)\,,
$$
$$ 
	[\,d^{\,j}_{1/2,-1/2}(z)\,]^2=[\,d^{\,j}_{1/2,1/2}(-z)\,]^2
\,.
$$
b) Bilinear terms with different $j$, the same $\lambda=1/2\,$, and $\mu=\pm1/2\,$:
$$ 
	d^{\,1/2}_{1/2,1/2}(z)\,d^{\,3/2}_{1/2,1/2}(z)=\frac12\,P_2(z)+\frac12\,P_1(z)\,,~~~
	d^{\,1/2}_{1/2,1/2}(z)\,d^{\,5/2}_{1/2,1/2}(z)=\frac12\,P_3(z)+\frac12\,P_2(z)\,,
$$
$$ 
	d^{\,3/2}_{1/2,1/2}(z)\,d^{\,5/2}_{1/2,1/2}(z)=\frac37\,P_4(z)+\frac15\,P_3(z)+
	\frac1{14}\,P_2(z)+\frac3{10}\,P_1(z)\,,
$$
$$ 
	d^{\,j_1}_{1/2,-1/2}(z)\,d^{\,j_2}_{1/2,-1/2}(z)=(-1)^{j_1+j_2-1}\,
	d^{\,j_1}_{1/2,1/2}(-z)\,d^{\,j_2}_{1/2,1/2}(-z)\,.
$$
c) Quadratic terms with $\lambda=3/2\,$:
$$ 
	[\,d^{\,3/2}_{3/2,1/2}(z)\,]^2=-\frac3{20}\,P_3(z)-\frac14\,P_2(z)
	+\frac3{20}\,P_1(z)+\frac14\,P_0(z)\,,
$$
$$ 
	[\,d^{\,5/2}_{3/2,1/2}(z)\,]^2=-\frac{25}{126}\,P_5(z)-\frac3{14}\,P_4(z)+
	\frac7{45}\,P_3(z)+\frac1{21}\,P_2(z)+\frac3{70}\,P_1(z)+\frac16\,P_0(z)\,,
$$
$$ 
	[\,d^{\,j}_{3/2,-1/2}(z)\,]^2=[\,d^{\,j}_{3/2,1/2}(-z)\,]^2\,.
$$
d) Bilinear terms with different $j$, the same $\lambda=3/2\,$, and $\mu=\pm1/2$:
$$ 
	d^{\,3/2}_{3/2,1/2}(z)\,d^{\,5/2}_{3/2,1/2}(z)=\sqrt{\frac32}\,\left[-\frac17\,
	P_4(z)-\frac15\,P_3(z)+\frac17\,P_2(z)+\frac15\,P_1(z)\right]\,,
$$
$$ 
	d^{\,j_1}_{3/2,-1/2}(z)\,d^{\,j_2}_{3/2,-1/2}(z)=(-1)^{j_1+j_2-1}\,
	d^{\,j_1}_{3/2,1/2}(-z)\,d^{\,j_2}_{3/2,1/2}(-z)\,.
$$

E) Expansions (for beam asymmetry) over associated Legendre functions:
$$  
	d^{\,1/2}_{1/2,-1/2}(z)\,d^{\,3/2}_{3/2,1/2}(z)
	=\frac{\sqrt{3}}{12}\,P_2^2(z)\,,~~~
	d^{\,1/2}_{1/2,-1/2}(z)\,d^{\,5/2}_{3/2,1/2}(z)=\frac{\sqrt{2}}{24}\,
	\left[P_3^2(z)-P_2^2(z)\right],
$$
$$ 
	d^{\,3/2}_{1/2,-1/2}(z)\,d^{\,3/2}_{3/2,1/2}(z)=
	\frac{\sqrt{3}}{8}\left[\frac15\,P_3^2(z)+\frac13\,P_2^2(z)\right],  
$$
$$ 
	d^{\,3/2}_{1/2,-1/2}(z)\,d^{\,5/2}_{3/2,1/2}(z)=
	\frac{\sqrt{2}}{8}\left[\frac17\,P_4^2(z)+\frac1{15}\,P_3^2(z)+\frac4{21}\,
	P_2^2(z)\right],
$$
$$  
	d^{\,5/2}_{1/2,-1/2}(z)\,d^{\,3/2}_{3/2,1/2}(z)=
	\frac{\sqrt{3}}{12}\left[\frac17\,P_4^2(z)+\frac15\,P_3^2(z)-\frac17\,
	P_2^2(z)\right],
$$
$$  
	d^{\,5/2}_{1/2,-1/2}(z)\,d^{\,5/2}_{3/2,1/2}(z)=
	\frac{\sqrt{2}}{16}\left[\frac{10}{63}\,P_5^2(z)+\frac2{21}\,P_4^2(z)+
	\frac4{45}\,P_3^2(z)+\frac47\,P_2^2(z)\right],
$$
$$ 
	d^{\,j_1}_{1/2,1/2}(z)\,d^{\,j_2}_{3/2,-1/2}(z)=(-1)^{j_1+j_2}\,
	d^{\,j_1}_{1/2,-1/2}(-z)\,d^{\,j_2}_{3/2,1/2}(-z)\,,~~~
	P_J^2(-z)=(-1)^J\,P_J^2(z)\,.
$$

\vspace{5mm}
\centerline{\bf Appendix~2.~Legendre coefficients}
\vspace{2mm}

A) Coefficients for cross section.\\
a) Even values of $J\,$:
$$ 
	\frac1{2N}\,A^{(\sigma)}_0(E)=2\left(\,|f_{1/2}^{1/2+}|^2+|f_{1/2}^{1/2-}|^2\right)+
	4\left(\,|f_{1/2}^{3/2+}|^2+|f_{1/2}^{3/2-}|^2\right)+6\left(\,|f_{1/2}^{5/2+}|^2+|f_{1/2}^{5/2-}|^2\right)
$$
$$ 
	+ 4\left(\,|f_{3/2}^{3/2+}|^2+|f_{3/2}^{3/2-}|^2\right)+6\left(\,|f_{3/2}^{5/2+}|^2+|f_{3/2}^{5/2-}|^2\right)\,;
	~~~~~~
$$
$$ 
	\frac1{2N}\,A^{(\sigma)}_2(E)=4\left(\,|f_{1/2}^{3/2+}|^2+|f_{1/2}^{3/2-}|^2\right)+
	\frac{48}7\left(\,|f_{1/2}^{5/2+}|^2+|f_{1/2}^{5/2-}|^2\right)~~~~~~~~~~~~~~~~~~~~~~~~~~~~$$
$$
	~~~~~~~~~~~~~~~~+8\,\mathrm{Re}\left(f_{1/2}^{1/2+*}f_{1/2}^{3/2+}+f_{1/2}^{1/2-*}f_{1/2}^{3/2-}\right)
	+12\,\mathrm{Re}\left(f_{1/2}^{1/2+*}f_{1/2}^{5/2+}+f_{1/2}^{1/2-*}f_{1/2}^{5/2-}\right)
$$
$$
	~~~~+\frac{24}7\,\mathrm{Re}\left(f_{1/2}^{3/2+*}f_{1/2}^{5/2+}+f_{1/2}^{3/2-*}f_{1/2}^{5/2-}\right)
	-4\left(\,|f_{3/2}^{3/2+}|^2+|f_{3/2}^{3/2-}|^2\right)
$$
$$
	~~~~~~~~~~~~+\frac{12}7\left(\,|f_{3/2}^{5/2+}|^2+|f_{3/2}^{5/2-}|^2\right)
	+\frac{24}7\sqrt{6}\,\mathrm{Re}\left(f_{3/2}^{3/2+*}f_{3/2}^{5/2+}+f_{3/2}^{3/2-*}f_{3/2}^{5/2-}\right)\,;
$$
$$ 
	\frac1{2N}\,A^{(\sigma)}_4(E)=\frac{36}7\left(\,|f_{1/2}^{5/2+}|^2+|f_{1/2}^{5/2-}|^2\right)
	+\frac{144}7\,\mathrm{Re}\left(f_{1/2}^{3/2+*}f_{1/2}^{5/2+}+f_{1/2}^{3/2-*}f_{1/2}^{5/2-}\right)~~~~~~~~
$$
$$
	~~~~~~~~~~~~~ -\frac{108}7\left(\,|f_{3/2}^{5/2+}|^2+|f_{3/2}^{5/2-}|^2\right)
	-\frac{24}7\sqrt{6}\,\mathrm{Re}\left(f_{3/2}^{3/2+*}f_{3/2}^{5/2+}+f_{3/2}^{3/2-*}f_{3/2}^{5/2-}\right)\,.
$$
b) Odd values of $J\,$:
$$
	\frac1{2N}\,A^{(\sigma)}_1(E)=4\,\mathrm{Re}\left(f_{1/2}^{1/2+*}f_{1/2}^{1/2-}\right)
	+\frac{8}{5}\,\mathrm{Re}\left(f_{1/2}^{3/2+*}f_{1/2}^{3/2-}\right)
	+\frac{36}{35}\,\mathrm{Re}\left(f_{1/2}^{5/2+*}f_{1/2}^{5/2-}\right)~~~~~~~~~~~~~~~~~~
$$
$$
	+8\,\mathrm{Re}\left(f_{1/2}^{1/2+*}f_{1/2}^{3/2-}+f_{1/2}^{1/2-*}f_{1/2}^{3/2+}\right)
+\frac{72}5\,\mathrm{Re}\left(f_{1/2}^{3/2+*}f_{1/2}^{5/2-}+f_{1/2}^{3/2-*}f_{1/2}^{5/2+}\right)~~~~~
$$
$$
	~~~~~~~~~~~~~~
	+\frac{24}{5}\,\mathrm{Re}\left(f_{3/2}^{3/2+*}f_{3/2}^{3/2-}\right)
	+\frac{108}{35}\,\mathrm{Re}\left(f_{3/2}^{5/2+*}f_{3/2}^{5/2-}\right)
	+\frac{24}5\sqrt{6}\,\mathrm{Re}\left(f_{3/2}^{3/2+*}f_{3/2}^{5/2-}+f_{3/2}^{3/2-*}f_{3/2}^{5/2+}\right)\,;
$$
$$
	\frac1{2N}\,A^{(\sigma)}_3(E)= \frac{72}{5}\,\mathrm{Re}\left(f_{1/2}^{3/2+*}f_{1/2}^{3/2-}\right)
	+\frac{32}{5}\,\mathrm{Re}\left(f_{1/2}^{5/2+*}f_{1/2}^{5/2-}\right)
	+12\,\mathrm{Re}\left(f_{1/2}^{1/2+*}f_{1/2}^{5/2-}+f_{1/2}^{1/2-*}f_{1/2}^{5/2+}\right)$$
$$
	~~~~~~~~~~~~~+\frac{48}5\,\mathrm{Re}\left(f_{1/2}^{3/2+*}f_{1/2}^{5/2-}+f_{1/2}^{3/2-*}f_{1/2}^{5/2+}\right)
	-\frac{24}{5}\,\mathrm{Re}\left(f_{3/2}^{3/2+*}f_{3/2}^{3/2-}\right)
	+\frac{56}{5}\,\mathrm{Re}\left(f_{3/2}^{5/2+*}f_{3/2}^{5/2-}\right)$$
$$
	-\frac{24}5 \sqrt{6}\,\mathrm{Re}\left(f_{3/2}^{3/2+*}f_{3/2}^{5/2-}+f_{3/2}^{3/2-*}f_{3/2}^{5/2+}\right)\,;
	~~~~~~~~~~~~~~~~~~~~
$$
$$
	\frac1{2N}\,A^{(\sigma)}_5(E)= \frac{200}{7}\,\mathrm{Re}\left(f_{1/2}^{5/2+*}f_{1/2}^{5/2-}\right)
	- \frac{100}{7}\,\mathrm{Re}\left(f_{3/2}^{5/2+*}f_{3/2}^{5/2-}\right)\,.
	~~~~~~~~~~~~~~~~~~~~~~~~~~~~~~~~~~~~~
$$
B) Coefficients for beam asymmetry.\\
a) Even values of $J\,$:\\
$$ 
	\frac1{4N}\,A^{(\Sigma)}_2(E)=
	\frac23\sqrt{3}\,\mathrm{Re}\left(f_{1/2}^{1/2-*}f_{3/2}^{3/2-}-f_{1/2}^{1/2+*}f_{3/2}^{3/2+}\right)
	-\frac{\sqrt{2}}2\,\mathrm{Re}\left(f_{1/2}^{1/2-*}f_{1/2}^{5/2-}-f_{1/2}^{1/2+*}f_{1/2}^{5/2+}\right)
$$
$$
	~~~~~~~~~~~~~~~~~~~ -\frac23\sqrt{3}\,\mathrm{Re}\left(f_{1/2}^{3/2-*}f_{3/2}^{3/2-}
	-f_{1/2}^{3/2+*}f_{3/2}^{3/2+}\right)
	-\frac27\sqrt{2}\,\mathrm{Re}\left(f_{1/2}^{3/2-*}f_{3/2}^{5/2-}-f_{1/2}^{3/2+*}f_{3/2}^{5/2+}\right)
$$
$$
	~~~~~~~~~~~~~~~~~~~~~~~~~~~ -\frac27\sqrt{3}\,\mathrm{Re}\left(f_{1/2}^{5/2-*}f_{3/2}^{3/2-}
	-f_{1/2}^{5/2+*}f_{3/2}^{3/2+}\right)
	+\frac97\sqrt{2}\,\mathrm{Re}\left(f_{1/2}^{5/2-*}f_{3/2}^{5/2-}-f_{1/2}^{5/2+*}f_{3/2}^{5/2+}\right)\,;
$$
$$ 
	\frac1{4N}\,A^{(\Sigma)}_4(E)=-\frac3{14}\sqrt{2}\,\mathrm{Re}\left(f_{1/2}^{3/2-*}f_{3/2}^{5/2-}
	-f_{1/2}^{3/2+*}f_{3/2}^{5/2+}\right)
	-\frac27\sqrt{3}\,\mathrm{Re}\left(f_{1/2}^{5/2-*}f_{3/2}^{3/2-}-f_{1/2}^{5/2+*}f_{3/2}^{3/2+}\right)
$$
$$
	+\frac3{14}\sqrt{2}\,\mathrm{Re}\left(f_{1/2}^{5/2-*}f_{3/2}^{5/2-}
	-f_{1/2}^{5/2+*}f_{3/2}^{5/2+}\right)\,.~~~~~~~~~~~~~~~~~~~~~~~~~~ 
$$
b) Odd values of $J\,$:
$$
	\frac1{4N}\,A^{(\Sigma)}_3(E)= \frac{\sqrt{2}}2\,\mathrm{Re}\left(f_{1/2}^{1/2-*}f_{3/2}^{5/2+}-f_{1/2}^{1/2+*}f_{3/2}^{5/2+}\right)
	-\frac25\sqrt{3}\,\mathrm{Re}\left(f_{1/2}^{3/2-*}f_{3/2}^{3/2+}-f_{1/2}^{3/2+*}f_{3/2}^{3/2-}\right)
$$
$$
	~~~~~~~~~~~~~~~~~~~~~ -\frac{\sqrt{2}}{10}\,\mathrm{Re}\left(f_{1/2}^{3/2-*}f_{1/2}^{5/2+}
	-f_{1/2}^{3/2+*}f_{1/2}^{5/2-}\right)
	+\frac25\sqrt{3}\,\mathrm{Re}\left(f_{1/2}^{5/2-*}f_{3/2}^{3/2+}-f_{1/2}^{5/2+*}f_{3/2}^{3/2-}\right)
$$
$$
	+\frac15\sqrt{2}\,\mathrm{Re}\left(f_{1/2}^{5/2-*}f_{3/2}^{5/2+}-f_{1/2}^{5/2+*}f_{3/2}^{5/2-}\right)\,;
	~~~~~~~~~~~~~~~~~~~~~~~~~
$$
$$
	\frac1{4N}\,A^{(\Sigma)}_5(E)=\frac5{14}\sqrt{2}\,\mathrm{Re}\left(f_{1/2}^{5/2-*}f_{3/2}^{5/2+}
	-f_{1/2}^{5/2+*}f_{3/2}^{5/2-}\right)\,.~~~~~~~~~~~~~~~~~~~~~~~~~~~~~~~~~~~~~~~~~~~
$$


\end{document}